\documentclass[aps,prb,twocolumn,showpacs,superscriptaddress,groupedaddress]{revtex4}  
 \usepackage{graphicx}  
\usepackage{dcolumn}   
\usepackage{bm}        
\usepackage{amssymb}   
\usepackage{epstopdf}
\usepackage{color}
\usepackage{dcolumn}
\usepackage{bm}

\def\m{\mathbf{m}}

\def\gapx{\lower 2pt \hbox{$\buildrel>\over{\scriptstyle{\sim}}$\ }}
\def\lapx{\lower 2pt \hbox{$\buildrel<\over{\scriptstyle{\sim}}$\ }}

\begin{document}
\title{Two holes in a two-dimensional quantum antiferromagnet:\\ A variational study based on  entangled-plaquette states}

\author{Fabio Mezzacapo}
\author{Adriano Angelone}
\author{Guido Pupillo}
\affiliation{icFRC, IPCMS (UMR 7504) and ISIS (UMR 7006), Universit\'e de Strasbourg and CNRS, 67000 Strasbourg, France}
\date{\today}	

\begin{abstract}															                  
We show that the entangled-plaquette variational ansatz can be adapted to  study  the two-dimensional $t-J$ model in the presence of two mobile holes. Specifically, we focus on a square lattice comprising up to $N=256$ sites in the parameter range $0.4\leq J/ t\leq2.0$. Ground state energies are  obtained via the optimization of a wave function  in which the weight of a given configuration is expressed in terms of variational coefficients associated with square and linear entangled plaquettes. Our estimates are in excellent agreement with  exact results available for the $N=16$ lattice. By extending our study to considerably larger systems we find, based on the analysis of the long distance tail of the probability of  finding two holes at spatial separation $r$, and on our computed two-hole binding energies,  the existence of a two-hole bound state for all the values of $J/t$ explored here. It is estimated that d-wave binding of the two holes  does not occur for  $J/t<J_c/t\simeq 0.19$.
 \end{abstract}


\maketitle

\section{Introduction} 

The theoretical investigation of the ground state properties of strongly correlated systems is one of the hardest problems in condensed matter physics. Many  relevant models lack an analytical solution, and the exact diagonalization (ED) of the hamiltonian matrix, while it can certainly offer useful insights, remains restricted to system sizes in general too small to provide a reliable description of the physical scenarios in the thermodynamic limit. In order to overcome this limitation a variety of numerical techniques have been developed, each of which has an optimal realm of applicability. For example, Quantum Monte Carlo (QMC) approaches\cite{qmc} provide essentially exact results for unfrustrated bosonic problems in any spatial dimension. However, QMC is hardly applicable without approximations to frustrated bosonic or fermionic systems, where  the so called ``sign problem" results in an exponential loss of accuracy of the results when decreasing temperature or increasing the number of particles.~\cite{sign} Conversely, variational approaches based on the optimization of a trial  wave function (WF) are sign-problem free; however, their accuracy ultimately depends on the choice and flexibility of the adopted ansatz for the WF. Recently, an impressive effort has been devoted to the development of tensor-network WF's able to describe strongly correlated systems in two spatial dimensions\cite{orus} (2D)  i.e., where the applicability of  Matrix Product States\cite{mps} and Density Matrix Renormalization Group\cite{dmrg} (DMRG) methods, extremely accurate in 1D, appears problematic.

One of the fundamental models used to characterize the behavior of strongly correlated electrons in 2D is the  $t-J$ model,\cite{and, rice} which is thought to provide an effective hamiltonian description of the basic features of superconducting copper oxides.  Key properties of the insulating copper-oxide planes at half-filling are reproduced by the spin-$1/2$  antiferromagnetic Heisenberg model i.e., the limiting case of the $t-J$ Hamiltonian in the absence of holes.\cite{man} The presence of mobile holes that may change the nature of the copper-oxide planes from insulating to superconducting is described in the $t-J$ model via an additional nearest-neighbor hopping term. The resulting Hamiltonian reads
\begin{equation}
H=-t\sum_{(i,j ),\sigma}(\overline{c}^+_{i,\sigma}\overline{c}_{j,\sigma}+h.c.)+J\sum_{(i,j)}(\mathbf{S}_i \cdot \mathbf{S}_j - \frac{1}{4}\hat{n}_i\hat {n}_j),
\label{eq:ham}
\end{equation}
where the  brackets restrict the sum to nearest neighbor sites of a square lattice comprising $L\times L=N$ sites. Here, $\overline{c}^+_{i,\sigma}=\hat{c}^+_{i,\sigma}(1- \hat{n}_{i,-\sigma})$ creates an electron with spin projection (e.g. along the $z$ axis) $\sigma$ on site $i$ excluding double occupancy, $\hat{c}^+_{i,\sigma}$ is the standard fermionic creation operator, while $\hat{n}_i=\hat{n}_{i,\sigma}+\hat{n}_{i,-\sigma}$  and  $\mathbf{S}_i$ are the number and spin-$\frac{1}{2}$  operator, respectively. In Eq.~(\ref{eq:ham})  $J>0$ is the antiferromagnetic coupling, and $t>0$ the hopping amplitude, taken in the following as energy unit.

Aside from its  physical interest related to its possible direct relevance to high-temperature superconductivity, the model Hamiltonian in Eq.~(\ref{eq:ham})  constitutes one of the  most challenging benchmarks to assess the accuracy of a given variational approach/WF.  For this problem ``exact" QMC techniques are applicable at half-filling,\cite{sand} where the $t-J$ model does  not  have fermionic character,  as well as to the static single-hole scenario. Accurate QMC strategies are also possible in the case of a single mobile hole.~\cite{mura} The addition of a second hole, however, introduces a severe sign problem that calls, in the QMC framework, for various, hardly controllable approximations and workarounds. A valid option to tackle the two-hole problem in (quasi-) 1D ladder geometries  is DMRG,\cite{weng} while in 2D the optimization of a suitable WF that allows for the investigation of system sizes larger than those treatable with ED likely represents  a preferable choice. In this framework the estimated ground state energy, as a strict upper bound of the actual value, constitutes a natural figure of merit to evaluate the accuracy of different ansatze.\\

In this paper we study the ground state of two holes in the $t-J$ model by using an entangled-plaquette WF.\cite{eps} Such a tensor-network-based WF,  founded on the variational family of the entangled-plaquette states (EPS), has been successfully employed to investigate different unfrustrated  and frustrated models providing results of comparable or better accuracy than those obtainable with alternative WF's or  techniques.\cite{cps, eps1, epstj1}
In the case of a single mobile hole,\cite{epstj1} for example, it provides estimates of ground state energy and hole spectral weight in excellent agreement with the most accurate results available in literature, based on QMC.\cite{mura}  Here, we show that an EPS WF including both square and linear plaquettes of limited sizes is able to faithfully describe the ground state of two holes in the $t-J$ model. The error on our estimates of the ground state energies for Eq. (\ref{eq:ham}) relative to the exact ones available for the $N=4\times4$ lattice is of the order or less than $0.1\%$ for all values of $J/t$ explored in this work. By considering square lattices of much larger size (i.e., up to $N=256$) we  show that binding of the two holes occurs for all of the analyzed  values of $J/t$; specifically, we find an exponential decay of the probability of finding two holes at distance $r$ in the large-$r$ limit and that the two-hole binding energy, although with an absolute value considerably smaller than the one of the system with $N=16$, stays negative in the thermodynamic limit. We estimate $J_c/t\simeq 0.19$ as the critical value below which the existence of a  bound state characterized by the $d_{x^2-y^2}$ symmetry, predicted by previous studies in our chosen parameter range,  is excluded. 

The accuracy of our findings for the two-hole $t-J$ model is a fundamental step towards the design of an EPS WF for the finite hole concentration scenario where the physics is still not completely understood. It is worth mentioning that relevant states proposed for the many-hole problem have a straightforward representation in terms of EPS\cite{cps} and essentially every WF  may  systematically be improved by taking advantage of the peculiar characteristics of the EPS ansatz (see below). 

The remainder of this paper is organized as follows: In the next section we discuss the EPS ansatz adopted in this work, also recalling the main properties of the general EPS WF. Then we present our results, and compare them with those obtained via alternative approaches. Finally we outline our conclusions pointing out possible extensions of the present work.

\section{Wave function}
\label{Sec:2}

\begin{figure}[t]
\centerline{\includegraphics[width = 1.0\columnwidth]{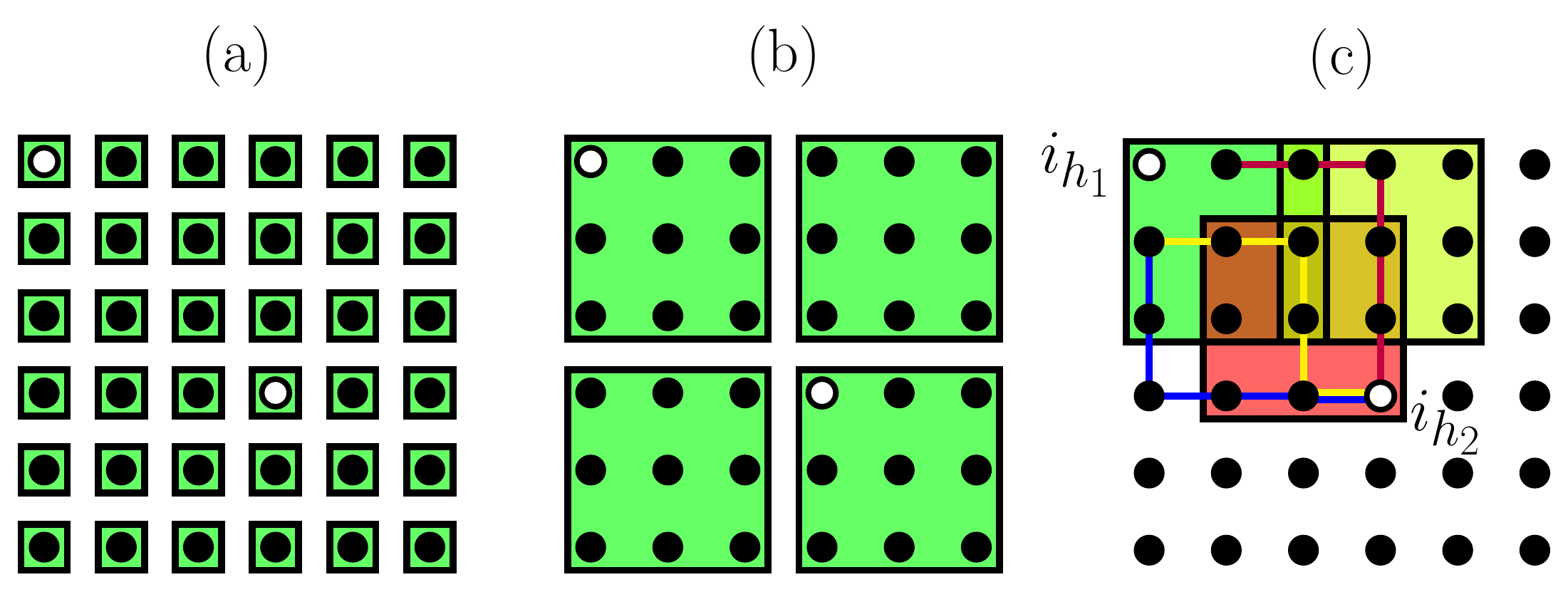}}
\caption{(color online). Graphic representation of various classes of plaquettes: (a) Single-site plaquettes; (b) $3\times3$ non overlapping plaquettes. Examples of $3\times 3$ entangled plaquettes and string-like plaquettes joining the hole in site $i_{h_2}$ with one of the nearest neighbor sites of the hole in  $i_{h_1}$ (see text) used to build the EPS WF in Eq.~(\ref{eq:wf}) are shown in panel (c).} 
\label{Fig:0}
\end{figure}

Let us consider an ensemble of $\mathcal{M}$ spin-1/2 particles on a lattice comprising $N$ no doubly occupied sites. The WF for such a system can be written as a weighted superposition of all possible configurations in the form: $|\Psi\rangle=\sum_{\m}W(\m)|\m\rangle$.~Here, $|\m\rangle=~|m_1,m_2,...m_N\rangle$, with $m_i=1$, ($-1$), or $0$ if site $i$ is occupied by a particle with ``up" (``down") spin projection along  an arbitrarily chosen axis, or empty. 
The general idea of the EPS  ansatz is to express the weight $W(\m)$ of a generic global configuration  $|\m\rangle$ in terms of variational coefficients in biunivocal correspondence with the configuration of  different groups of sites i.e., plaquettes. The simplest (non entangled) plaquette ansatz consists of choosing $W(\m)=\prod_{P=1}^NC_P^{m_{i_{1,P}}}$ where  $C_P^{m_{i_{1,P}}}$ are variational coefficients associated with the configuration of the single site [see Fig.~\ref{Fig:0}(a)], labeled by $m_{i_{1,P}}$, of the $P_{th}$ plaquette.
This choice results in a mean-field-like WF where correlations are neglected. However, they can be promptly incorporated in the ansatz by increasing the plaquette size. While in the case of non overlapping plaquettes [Fig.~\ref{Fig:0}(b)] correlations are well described for distances of the order of the plaquette size, a reliable description of long range correlations is obtainable, even with relatively small plaquettes, when the latter overlap (i.e., are entangled). Clearly, any EPS ansatz is a legitimate variational choice regardless of the size of the plaquette used. In other words one can adopt a given plaquette size and provide variational estimates with an accuracy related to the given dimension of the plaquettes. This is exactly as in any variational calculation based on different WF?s, where the accuracy is related to the chosen variational ansatz.  Furthermore, the EPS WF is systematically improvable  by enlarging the size of the plaquettes and/or by including plaquettes of various shapes correlating specific groups of sites, being exact in the limit of a single plaquette as large as the system. 

Our adopted EPS WF for the study of the ground state properties of the Hamiltonian (\ref{eq:ham}) on an $N$-site square lattice with periodic boundary conditions, in the presence of two mobile holes, is
\begin{equation}
|\Psi\rangle=\sum_{\m,S}(-1)^{\mathcal{L}(\m)+\mathcal{F}(\m_S)}\prod_PC^{\m_P}_PC_S^{\m_S}|\m\rangle,
\label{eq:wf}
\end{equation}
where $\m$ refers to a configuration with $S^{TOT}_z=0$  comprising $N-2$ electrons and $2$ holes at sites $i_{h_1}$ and $i_{h_2}$ (with $i_{h_1}<i_{h_2}$). The above ansatz includes two classes of plaquettes characterized by index $P$ and $S$, respectively. Specifically, we consider $N$ square plaquettes comprising $l$  sites, where their configuration, for a given plaquette, is labeled via $m_{i_{1,P}},...,m_{i_{l,P}}$ and linear string-like plaquettes joining sites $i_{h_2}$ and $i_{h_1}$ comprising $l'$ sites so that $i_{1,S}\equiv i_{h_2}$ and $i_{l',S}$ is a nearest neighbor of $i_{h_1}$. Examples of plaquettes belonging to both classes are illustrated in Fig.~\ref{Fig:0}(c). In Eq.~(\ref{eq:wf}), $\mathcal{L}(\m)=N^{A}_\downarrow+\sum_{i \in A, j> i}\widehat{n}^h_i\widehat{n}^h_j$,  with $N^{A}_\downarrow$ the number of down spins in one of the two sublattices of the square lattices and $\widehat{n}^h_i $ the hole-number operator at site $i$;  similarly $\mathcal{F}(\m_S)=N^{S}_\downarrow+g(i_{h_1}, i_{l',S})$ where the first term on the right-hand side counts the number of down spins comprised in the $S_{th}$ string-like plaquette and $g(i_{h_1}, i_{l',S})=1$ $(0)$ if the distance between $i_{h_1}$ and $i_{l',S}$ is $\pm\widehat{\mathbf{y}}$ $(\pm\widehat{\mathbf{x}})$. The resulting phase factor $(-1)^{\mathcal{L}(\m)+\mathcal{F}(\m_S)}$, reduces at half-filling to the exact Marshall sign rule,\cite{mar} and, for the present study, is found, in our explored parameter range, to improve the optimization of the wave function favoring the emergence of the ground state properties of the system\\

In our calculations we set $l=9$ corresponding to $3\times3$ plaquettes and consider string plaquettes comprising up to $l'=l$ sites. A null weight has been assigned to system configurations in which holes are connected by longer strings. We carry out independent optimizations of the state in Eq.~(\ref{eq:wf}) for each lattice size and value of $J/t$ considered here via the variational Monte Carlo algorithm described in Ref.~[\onlinecite{eps}] and use the same numerical approach to estimate the observables of our interest. In particular, for $0.4\leq J/t \leq 2.0$ we compute (i) the two-hole ground state energy defined as $\delta E_2/t=(E_2-E_0)/t$, where $E_{2}$ ($E_0$) are the ground state energies of model~(\ref{eq:ham}) with two holes (at half filling), (ii) the probability distribution $P(r)=\sum_{i<j}\widehat{n}^h_i\widehat{n}^h_j\delta(r_{ij}-r)$ of finding the two holes at distance $r$, as well as (iii) the two-hole binding energy $\Delta/t=\delta E_2/t-2\delta E_1/t$ where the one-hole ground state energy $\delta E_1/t$ has been estimated by means of the EPS ansatz based on $3\times3$ plaquettes proposed by one of us in Ref.~[\onlinecite{epstj1}]. It has to be stressed  that with our chosen dimension of the plaquettes  we obtain remarkable agreement with ED calculations\cite{barnes, dag, poi} for both the single- and the two-hole problems; similarly, on large lattices our estimates of both the single- and the two-holes ground state energies are in extremely good agreement with the most accurate results available in literature\cite{mura, mas} (see Sec.~\ref{results}). This is an important point since a consistent increase of the plaquette size e.g., by considering square plaquettes of 16 sites, if doable, would be extremely expensive from a computational point of view due to the dimension (i.e.,~3) of the local Hilbert space of the $t-J$ model.  Our findings for $P(r)$ obtained for lattices of up to $N=256$ sites, that is, much larger than those treatable with exact methods, demonstrate the existence of a two-hole bound state for any value of $J/t$ considered here. Estimates of the two-hole binding energy extrapolated to the thermodynamic limit and for values of $J/t$ lower than $0.4$ suggest that a two-hole bound state does not exist with the same symmetry characteristic of the range  of $J/t$ values explored in this work for $J/t \lesssim 0.19$.
\section{RESULTS}
\label{results}
\begin{figure}[b]
\centerline{\includegraphics[width = 1.0\columnwidth]{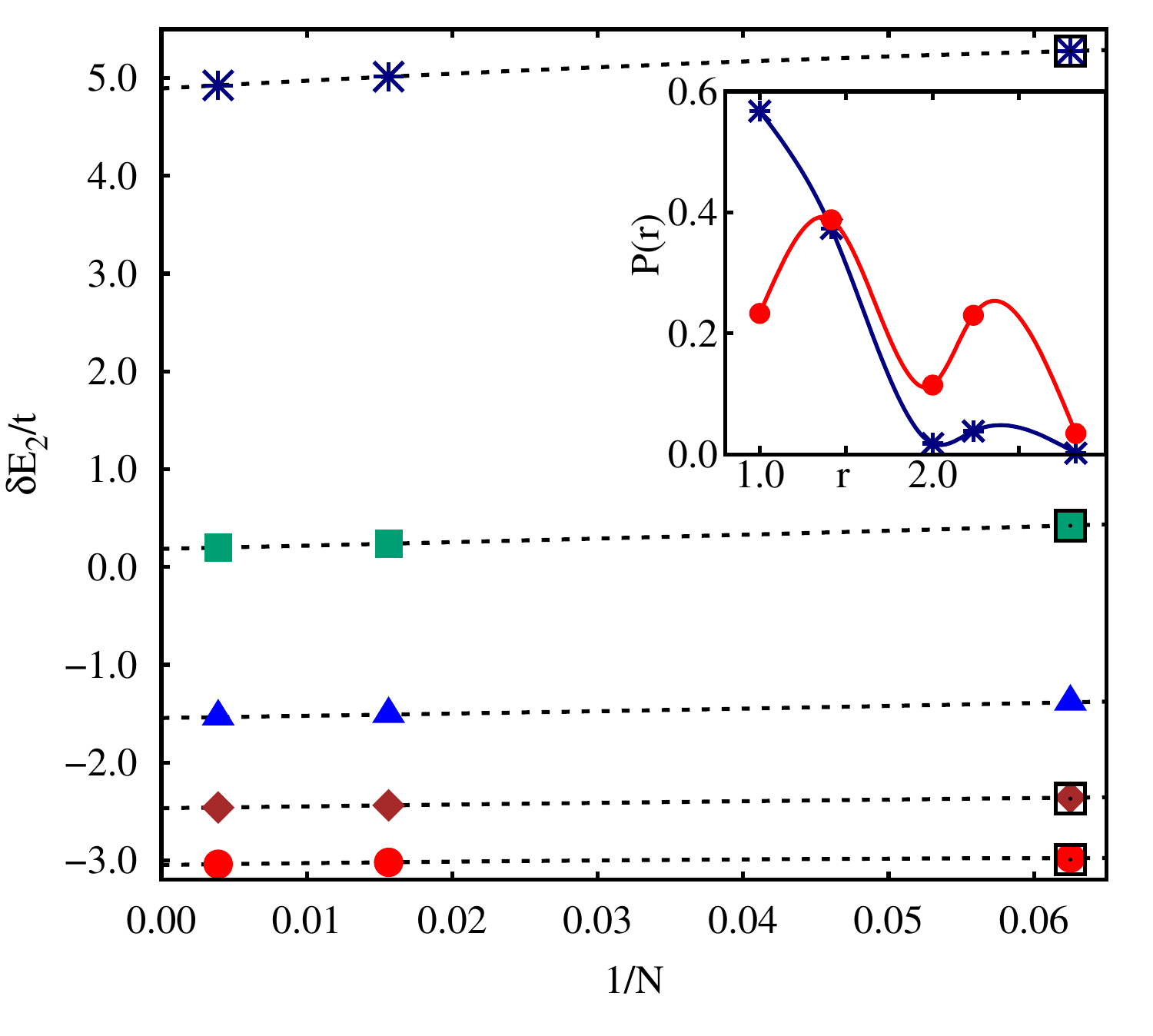}}
\caption{(color online). Two-hole ground state energy $\delta E_2/t$ of model Eq.~(\ref{eq:ham}) as a function of the lattice size $N$.  Estimates are obtained with the EPS ansatz in Eq.~\ref{eq:wf}. Values of $J/t$ are 2.0 (stars), 1.0 (squares), 1/1.5 (triangles), 0.5 (diamonds) and 0.4 (circles). Error bars are smaller than the symbol size. Exact results\cite{barnes, dag} available for the $4\times4$ lattice are also shown (empty squares) for comparison. The dotted lines are polynomial in the inverse system size fitting functions to numerical data. Inset: Probability $P(r)$ of finding two holes at distance $r$ on a $4\times4$ lattice; same symbols correspond to the same values of $J/t$ in the main panel, solid lines are guides to the eye. Distances are in units of the lattice constant.} 
\label{Fig:1}
\end{figure}
\begin{figure}[t]
\centerline{\includegraphics[width = 1.0\columnwidth]{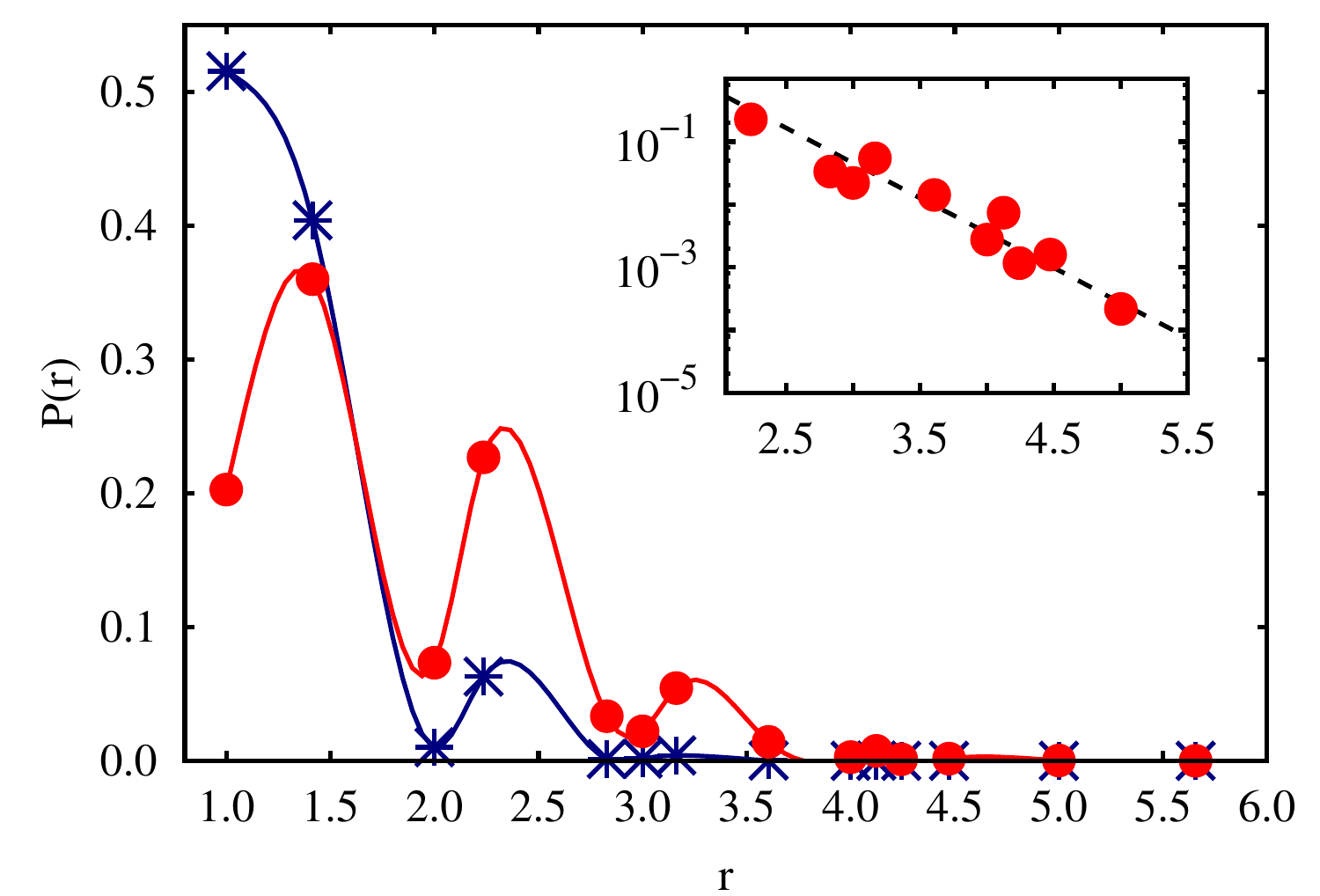}}
\caption{(color online). Probability $P(r)$ of finding two holes at distance $r$ on the $8\times8$ lattice. Estimates are obtained with the EPS ansatz in Eq.~\ref{eq:wf}. Values of $J/t$ are 2.0 (stars) and 0.4 (circles). Error bars are smaller than the symbol size; solid lines are guides to the eye. Inset: large distance decay of $P(r)$ for $J/t=0.4$; also shown the exponential (see text) fitting function  adopted to  describe our numerical data (dashed line). Distances are in units of the lattice constant.} 
\label{Fig:2}
\end{figure}
While for small system sizes (i.e., up to $N\simeq 16$) it is possible to describe the ground state properties of model~(\ref{eq:ham}) essentially exactly by means of an EPS WF based on a single plaquette that correlates all the lattice sites, such a choice is not a viable option for larger lattices. Our variational state in Eq.~(\ref{eq:wf}), where plaquettes comprising a limited number of sites are used,  provides accurate energy upper bounds for the lattice with $N=16$ and is applicable to considerably larger lattice sizes using standard computational resources. For example, on the $4\times4$ square lattice we find, at $J/t=1.0$, $E_2=-18.8007(1)t$ which compares extremely well with the exact result\cite{barnes} $E_2^{ex}=-18.8061t$. The resulting EPS two-hole  ground state energy is $\delta E_2/t( J/t=1.0)=0.4246(1)$, which has to be compared with $\delta E_2^{ex}/t( J/t=1.0)=0.4223$. It is interesting to contrast our results with those obtained by means of a Green's function Monte Carlo (GFMC) approach based on the extrapolation of transient energy estimates generated by the GFMC algorithm starting from a suitable initial state. For the two-hole $t-J$ model, the GFMC technique is affected by the fermionic sign problem and the mentioned extrapolation can be performed by using just a few transient estimates before the occurrence of an uncontrolled growth of the statistical uncertainty ultimately due to sign instability. 
Consequently, the choice of the initial state is crucial in the case of GFMC as it has to produce reliable estimates in a limited number of algorithm iterations. Although for $J/t=1.0$ this procedure gives an extrapolated value  $\delta E^{GFMC}_2/t (J/t=1.0) =0.42(1)$, in agreement with our EPS result, we note that the GFMC  zero-th, variational, iteration based on the initial WF provides a two-hole ground state energy more than $3$ times larger. This demonstrates that our EPS ansatz is much more accurate than the initial variational state adopted in Ref.~[\onlinecite{mas}] and, more importantly, suggests our WF as a nearly optimal one to start a GFMC numerical scheme consisting of few iterations. The latter, aside from the above mentioned possibility of adding variational flexibility to a general EPS WF by including larger plaquettes, constitutes a further opportunity to improve  numerical estimates.

Figure \ref{Fig:1} shows EPS results for the two-hole ground state energy $\delta E_2/t$  as a function of the system size $N$ and various values of $J/t$. The relative error of our numerical estimates with respect to the exact results obtainable for the $N=16$ lattice (i.e., the smallest considered here) is of the order of $0.5\%$ or less regardless of the $J/t$ value. On larger lattices our two-hole ground state energies compare extremely well with GFMC ones; at $J/t=1.0$, for example,  our estimated value for the $8\times8$ system is $0.238(2)$ in numerical agreement, taking into account the quoted error bars,  with the GFMC result  i.e., $0.26(2)$.\cite{mas} By means of a simple extrapolation of our data to the thermodynamic limit based on a polynomial expansion in powers of $1/N$ (dashed lines in figures) we find that the two-hole ground state energy monotonically decreases with decreasing $J/t$ being e.g.,  $\delta E^{N=\infty}_2/t(J/t=1.0)\simeq 0.185$ and   $\delta E^{N=\infty}_2/t(J/t=0.4)\simeq -3.05$. Our extrapolated results are in substantial agreement with the estimates for the largest lattice size studied in this work (i.e., $N=256$) pointing out how the EPS ansatz allows, for the model of our interest, to  investigate lattices large enough to provide a good approximation of the physics emerging in the thermodynamic limit. The probability $P(r)$ of finding the two holes at distance $r$ on the $4\times4$ lattice for chosen values of $J/t=2.0$  and $0.4$  is plotted in the inset of Fig.~\ref{Fig:1}. This quantity displays an oscillating behavior with a global maximum at $r=1$ for  $J/t=2.0$. For lower $J/t$ the position of such a maximum shifts to $r=\sqrt{2}$ and $P(r)$ at larger $r$ increases, signaling an enhanced propensity of the two holes to reside on distant lattice sites. This may possibly result for larger system sizes in an ``unbound" two-hole ground state. Conversely, if the two holes form a bound state $P(r)$ is expected to feature an exponential decay at large distances.\cite{polaron}

\begin{figure}[t]
\centerline{\includegraphics[width = 1.0\columnwidth]{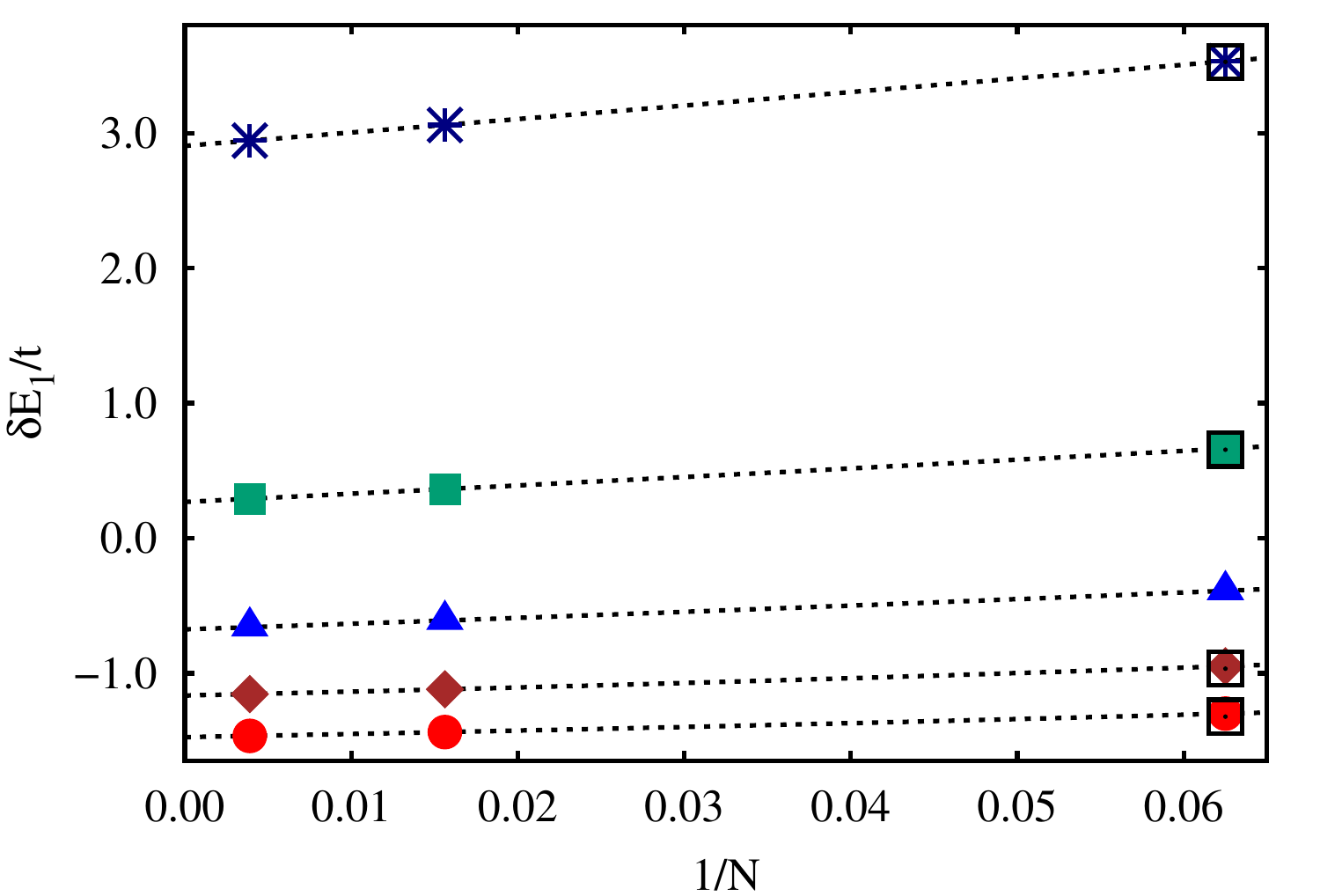}}
\caption{(color online). Single-hole ground state energy $\delta E_1/t$ of model Eq.~(\ref{eq:ham}) as a function of the lattice size $N$.  Estimates are obtained with the EPS ansatz used in Ref.~[\onlinecite{epstj1}]. Values of $J/t$ are 2.0 (stars), 1.0 (squares), 1/1.5 (triangles), 0.5 (diamonds) and 0.4 (circles). Error bars are smaller than the symbol size. The dotted lines are polynomial in the inverse system size fitting functions to numerical data. Exact results available for the $4\times4$ lattice are also shown (empty squares) for comparison.} 
\label{Fig:3}
\end{figure}

Figure \ref{Fig:2} shows estimates of $P(r)$ on a lattice of $N=64$ sites. Although the qualitative behavior of the two-hole distribution function is similar to that found for $N=16$ here, as expected, holes are more separated on average. The smaller is the value of $J/t$, the larger is their tendency to increase their relative distance. However, for large $r$ our data are well described by the simple functional form $P(r) \sim e^{-r/\xi}$ where, for $J/t=0.4$ (see inset), we estimate $\xi \sim 0.4$.
By increasing the lattice size to $N=256$ the value of $\xi$ stays essentially unchanged. 
On the basis of this analysis we can conclude that the two holes form a bound state for all the values $0.4\leq J/t \leq 2.0$ examined in our study.

Quantitative information about the two-hole bound state are obtainable by computing the binding energy $\Delta/t$ defined in Sec.~\ref{Sec:2}. A negative value of this quantity signals the existence of the bound state. In order to estimate $\Delta/t$, both the two- and the single-hole ground state energies are needed. The single-hole ground state energy is plotted as a function of the system size in Fig.~\ref{Fig:3}, for several values of $J/t$. The binding energy resulting from the combination of data in Fig. \ref{Fig:1} and Fig. \ref{Fig:3} displays a marked dependence on the system size as well as on the values of $J/t$. For example, for $N=256$ we find $\Delta/t(J/t=0.4)=-0.111(3)$, a value in agreement with the GFMC estimate of $-0.12(4)$, approximately $3$ times higher than that for the $4\times4$ lattice. On the other hand, on a $16\times16$ lattice when $J/t$ increases from $0.4$ to $1.0$, the two-hole binding energy decreases down to $\sim -0.39$. 

\begin{figure}[t]
\centerline{\includegraphics[width = 1.0\columnwidth]{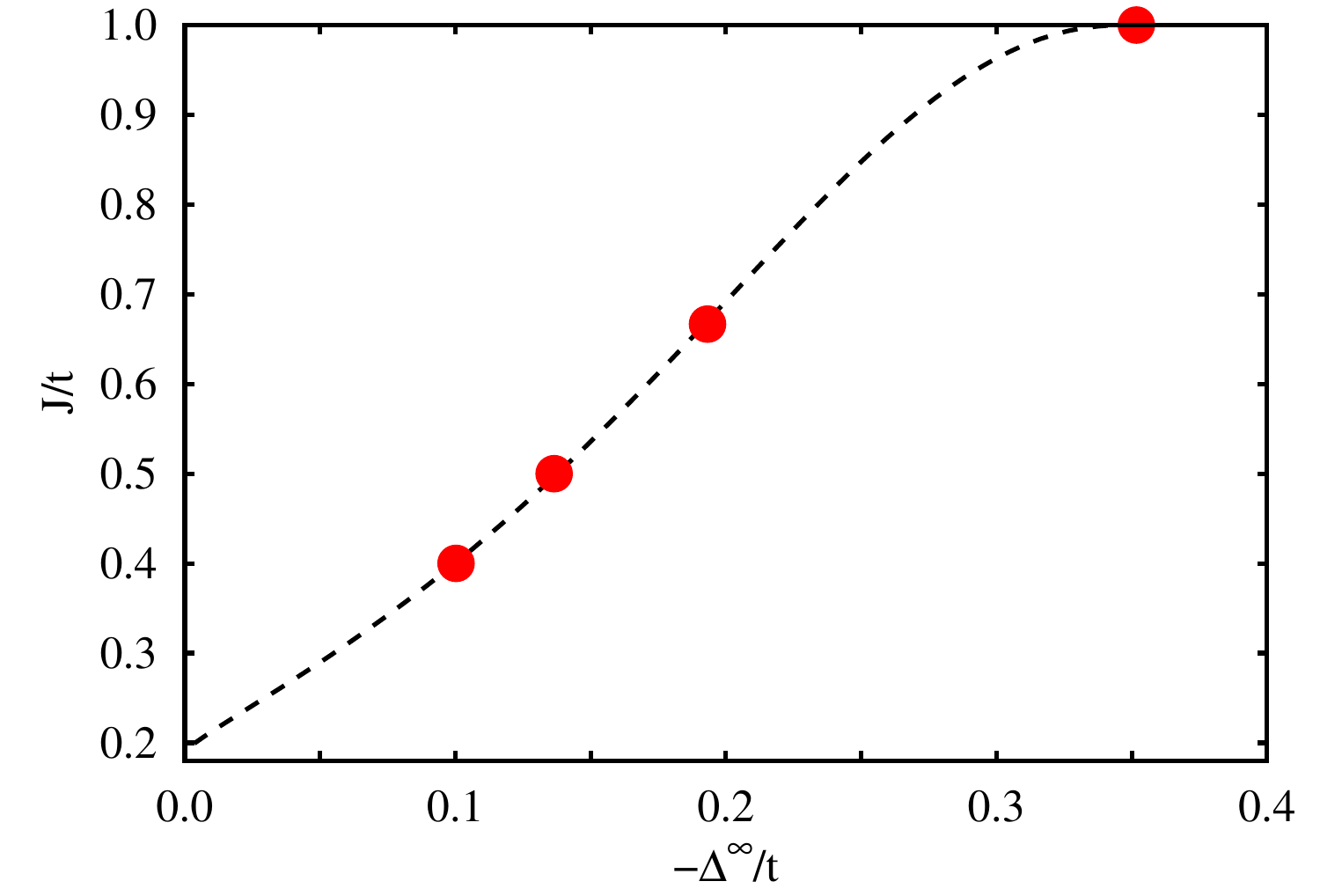}}
\caption{(color online). $J/t$ versus opposite binding energy extrapolated to the thermodynamic limit  $-\Delta^{\infty}/t$. The dashed line is a fitting function to our numerical estimates (see text).} 
\label{Fig:4}
\end{figure}

Values of the binding energy extrapolated to the thermodynamic limit are plotted in Fig. \ref{Fig:4}. Specifically, for each value of $J/t$, $\Delta^{\infty}/t=\delta E^{\infty}_2/t-2\delta E^{\infty}_1/t$ is computed via the corresponding extrapolations of the two- and single-hole ground state energies (see dashed lines in Figs. \ref{Fig:1} and \ref{Fig:3}, respectively). By assuming, as in Ref.~[\onlinecite{mas}], the functional dependence
$t/J= \mathcal{G}(x=\Delta^{\infty}/t)=t/J_c[1-\lambda x \ln (x/\epsilon)]$, we estimate the critical value $J_c\simeq 0.19 t$ at which the two-hole binding energy extrapolated to the thermodynamic limit reaches zero. This estimate, in agreement with that obtained in the case of the $16\times 16$ lattice, indicates that for $J\lesssim J_c$ a bound state of two holes, if present, is characterized by a symmetry different from that (i.e., d-wave) predicted by several studies in the parameter range of Fig.~\ref{Fig:4}. Indeed,  a change in the symmetry of the bound state should occur for $J/t\lesssim 0.18$\cite{b1} (or~$0.15$).\cite{b2}\\

\section{Conclusions and Perspectives}

We have shown that the entangled-plaquette variational ansatz can be applied to study the ground state properties of two mobile holes in a two-dimensional quantum antiferromagnet for lattice sizes considerably larger than those treatable with exact approaches. Obtained energy estimates are in remarkable agreement with exact results on a $N=16$ lattice. We have extended our analysis to a maximum system size of $N=256$, demonstrating the existence of a two-hole bound state for all the values of $J/t$ explored here. An extrapolation of our estimated two-hole binding energy in the large $N$ limit to low values of $J/t$ results in a critical $J_c\simeq0.19t$ below which a bound state with d-wave symmetry is not expected. Including e.g., the p-wave symmetry in the EPS ansatz to investigate the existence of a different two-hole bound state in the ground state for  $J/t \lesssim 0.19$ as well as studying the dependence of the physical properties discussed here on the presence of a next-nearest-neighbor hopping term in Eq. (\ref{eq:ham}) are possible interesting extensions of the present work.

Furthermore, although specific QMC approaches can still be adopted for the two-hole $t-J$ model at the price of a large error bar on the resulting estimates, in the finite hole density scenario, where the physical picture remains under debate,\cite{c1,b,c2}  their applicability is unfeasible due to an even heavier sign problem. In such a case  the EPS ansatz, in the framework of a simple, by definition sign-problem free, variational approach  (e.g., that employed here)  may constitute a viable option either as a generalization of Eq. (\ref{eq:wf}) or as a systematic route to improve relevant wave functions.

\section*{Acknowledgments}
This research is supported by the European Commission via ERC-St Grant ‘‘ColdSIM’’
(No. 307688). We acknowledge additional partial support from EOARD,  H2020 FET Proactive project RySQ (grant N. 640378),  ANR-FWF via "BLUSHIELD", and UdS via Labex NIE and IdEX, computing time at the HPC-UdS.

\end{document}